\documentclass[conference]{IEEEtran}
\IEEEoverridecommandlockouts
\usepackage[T1]{fontenc}
\usepackage[utf8]{inputenc}
\usepackage{graphicx}

\begin{document}

\title{A Software-Defined Testbed for Quantifying Deauthentication Resilience in Modern Wi-Fi Networks\\

\thanks{”This work has been submitted to the IEEE for possible publication. Copyright may be transferred without notice, after which this version may no longer be accessible.”}
}

\author{
\IEEEauthorblockN{Alex Carbajal and Asma Jodeiri Akbarfam}
\IEEEauthorblockA{School of Engineering and Applied Sciences, Washington State University Tri-Cities\\
alexander.carbajal@wsu.edu, asma.akbarfam@wsu.edu}
}

\maketitle

\begin{abstract}

Wi-Fi deauthentication attacks remain a practical denial-of-service (DoS) threat by exploiting unprotected management frames to disrupt client connectivity. In this work, we introduce a software-defined testbed to measure Wi-Fi resilience to deauthentication attacks. We experimentally evaluate five wireless security configurations: open networks, WPA1, WPA2 without Protected Management Frames (PMF), WPA2 with PMF, and WPA3. Using controlled experiments, we measure client disconnection rates, packet injection volume, and time-to-disruption under each configuration. Packet-level behavior is analyzed using standard wireless auditing tools. Open networks, WPA1, and WPA2 without PMF proved entirely vulnerable to deauthentication, while no successful attacks were observed for WPA2 with PMF or WPA3 under tested conditions. These findings confirm the effectiveness of management-frame protection and highlight the continued risk posed by legacy or misconfigured wireless deployments.

\end{abstract}

\section{Introduction}
\label{sec:introduction}

Wireless connectivity has become a central component of modern computing infrastructures, with many networks today supporting more wireless connections than wired ones \cite{rikitianskaia}. Wi-Fi technologies enable ubiquitous internet access without physical cabling, but this convenience also introduces an expanded attack surface that adversaries can exploit \cite{gierszewski-matuśkiewicz}. In practice, Wi-Fi networks expose three primary points of vulnerability: the client device, the access point, and the wireless medium \cite{beard-stallings}. This work experimentally examines vulnerabilities across all three through the lens of deauthentication-based denial-of-service \cite{zanna-radcliffe-kumar} attacks.

Wi-Fi deauthentication is a denial-of-service technique that targets management frames to forcibly disconnect legitimate clients from access points, interrupting communication and degrading service availability \cite{schepers-ranganathan-vanhoef}. In addition to disruption, such attacks enable follow-on threats including traffic inference, session interruption, and credential capture during client reconnection attempts \cite{zanna-radcliffe-kumar}. While the underlying attack mechanism is well understood, real-world susceptibility varies significantly depending on the security protocol in use and the configuration of management-frame protection mechanisms \cite{faíscas}.

In this study, we quantify the resilience of contemporary Wi-Fi networks to classical deauthentication injection attacks using a reproducible, software-defined access-point testbed. Classical deauthentication attacks abuse unsecured Wi‑Fi management frames by sending spoofed disassociate packets from an attacker that is within monitoring and injection range of the network \cite{schepers-ranganathan-vanhoef}. We evaluate five representative configurations: open networks, WPA1, WPA2 without Protected Management Frames (PMF), WPA2 with PMF, and WPA3. Experimental measurements include client disconnection rates, packet injection volumes, and time-to-disruption under controlled attack conditions. The objective is to assess the operational limits of legacy deauthentication techniques under contemporary security configurations rather than to provide a procedural attack guide.

Although deauthentication attacks have been documented in the literature \cite{schepers-ranganathan-vanhoef}, \cite{chatzoglou-kambourakis-kolias}, systematic measurement across modern security configurations remains limited, particularly in software-defined environments that closely mirror dynamic real-world deployments. By performing controlled experimentation across multiple configurations within a uniform framework, this work provides a quantitative comparison of deauthentication resilience and highlights the defensive impact of management-frame protection in contemporary networks.

Motivated by the increasing dependence on wireless infrastructure in public, private, and operational settings \cite{rikitianskaia}, this work aims to clarify the protection boundaries of modern Wi-Fi protocols against denial-of-service threats. As Wi-Fi networks continue to expand into safety-critical and high-reliability domains \cite{gierszewski-matuśkiewicz}, understanding the practical limits of security mechanisms such as PMF and WPA3 is essential to informed deployment and risk management.

Unlike existing studies that focus on isolated configurations or vendor-specific platforms \cite{schepers-ranganathan-vanhoef}, \cite{chatzoglou-kambourakis-kolias}, this work makes the following contributions:
\begin{itemize}
    \item The design of an automated, software-defined testbed that allows for the user-friendly and reproducible evaluation of Wi-Fi security protocols. This system provides critical insights that directly inform how to securely deploy modern wireless networks.
    \item A quantitative comparison of deauthentication resilience across five commonly deployed configurations under identical conditions.
    \item Empirical isolation of the impact of Protected Management Frames (PMF) on mitigating classical deauthentication attacks.
\end{itemize}
Together, these contributions provide a practical benchmark for assessing operational Wi-Fi resilience.

\section{Background}
\label{sec:background}

This work investigates the resilience of Wi-Fi networks to classical deauthentication attacks under different security configurations using a controlled, software-defined testbed.

The experimental setup relies on a software-based wireless access point (WAP) implemented on an Arch Linux system using wireless network interface controllers (WNICs) \cite{gast-wireless}. Software access points are widely used for protocol testing and experimental validation, enabling fine-grained control over authentication behavior and management-frame handling \cite{hostapd}. The underlying wireless communication follows the IEEE 802.11 standard \cite{stacey}, with modern deployments employing advanced versions such as 802.11ax \cite{ieee-wifi-6} and 802.11be \cite{ieee-wifi-7} for improved performance and capacity.

Security evaluation is performed across open networks, WPA1, WPA2 without PMF, WPA2 with PMF, and WPA3. After the weaknesses of WEP were revealed \cite{wep}, Wi-Fi Protected Access (WPA) was introduced to provide encryption and authentication for wireless communication \cite{reddy-srikanth-wpa}. WPA-based deployments commonly operate in either pre-shared key (WPA-PSK) modes or the SAE mode introduced with WPA3, both of which are supported in the testbed.

A central mechanism evaluated in this study is Protected Management Frames (PMF), which protects authentication and association frames from forgery and replay \cite{ieee-pmf}. PMF is mandatory in WPA3 and optional in WPA2. When enabled, it blocks classical deauthentication injection attacks by enforcing cryptographic integrity on management traffic.

Client devices in the experimental network are uniquely identified using MAC addresses \cite{ieee-mac} and are assigned IP addresses dynamically through DHCP \cite{lemon-sommerfeld-dhcp}. Name resolution is provided using DNS services \cite{mockapetris-dns}, while external connectivity is enabled through Network Address Translation (NAT) \cite{srisuresh-egevang-nat}. These mechanisms collectively support a realistic operational network environment.

This configuration establishes the technical basis for the experimental measurements and comparative evaluation presented in this paper.

\section{Related Work}
\label{sec:related-work}

Wi-Fi deauthentication attacks typically involve a multitude of software tools that aid in the practical implementation of an attack. One of these tools is \textit{Aircrack-ng}, which is a suite of tools with the purpose of auditing the security of Wi-Fi networks. Another tool, \textit{mdk4}, allows a wireless device to inject packets into a wireless network and exploit Wi-Fi vulnerabilities. This includes dissociate packets that force clients connected to a wireless network to forcibly deauthenticate. The official documentation of \textit{Aircrack-ng} provides a tutorial and examples that demonstrate how one would perform such an attack \cite{aircrack-ng}.

An academic study by E. Chatzoglou \textit{et al.} \cite{chatzoglou-kambourakis-kolias} demonstrates the security of modern Wi-Fi technology by performing a DoS attack on a WPA3-SAE secured wireless network. Normally, Wi-Fi DoS attacks are performed on networks that are open, using WEP or WPA1, or using WPA2 without PMF enabled. WPA3-SAE requires PMF to be enabled in order to function, and their work demonstrates that denial-of-service still remains possible under alternative attack models distinct from classical deauthentication injection. Additionally, in their research paper involving countermeasures for Wi-Fi deauthentication attacks, D. Schepers, A. Ranganathan, and M. Vanhoef \cite{schepers-ranganathan-vanhoef} examine similar vulnerabilities. They come to the conclusion that Management Frame Protection (MFP) is not sufficient to protect a Wi-Fi network against deauthentication attacks \cite{schepers-ranganathan-vanhoef}.

In contrast to prior work exploring advanced DoS vectors against WPA3-SAE, this study focuses on classical deauthentication injection under misconfigured and legacy security settings. We introduce a reproducible, software-defined testbed for controlled evaluation of deauthentication resilience across multiple Wi-Fi security configurations. In contrast to tutorial-style demonstrations or single-configuration analyses, our approach enables direct comparison under identical conditions while emphasizing empirical measurement and operational impact. Rather than serving as a procedural guide, this work focuses on characterizing resilience boundaries and validating the effectiveness of management-frame protection in contemporary deployments.

\section{System Design / Methodology}
\label{sec:system-design}

\begin{figure}[ht]
    \centering
    \fbox{\includegraphics[width=0.9\linewidth]{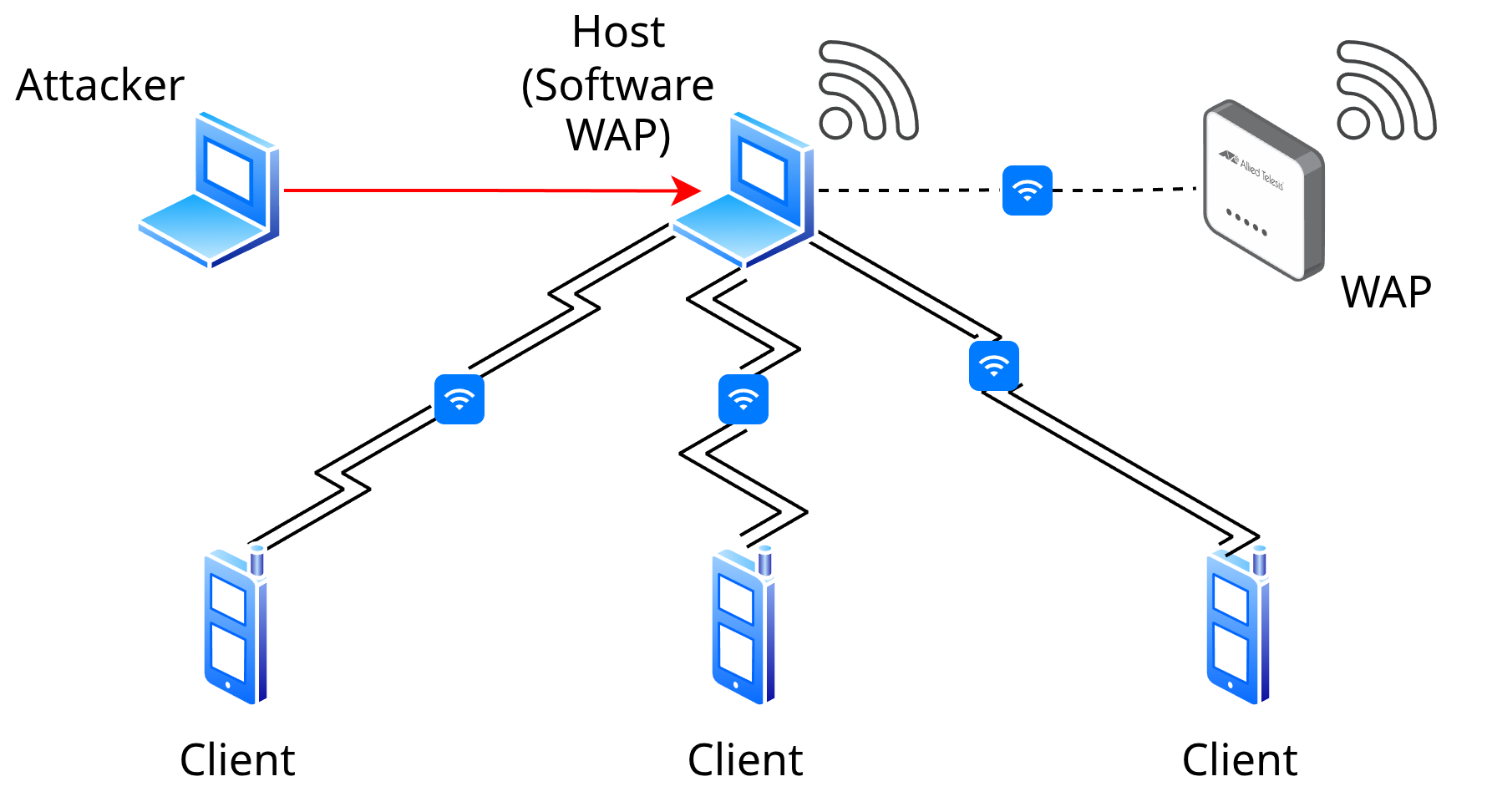}}
    \caption{Client-Host-Attacker Network Topology}
    \label{fig:network-topology}
\end{figure}

One Arch Linux system will serve as the host machine, acting as the source of the software WAP. A second Arch Linux system will serve as the attacker, performing the deauthentication attack on the software WAP. Each system contains a WNIC capable of AP mode, monitor mode, and packet injection. Clients will connect directly to the host using Wi-Fi. To ensure connectivity to the internet for clients, the host will be connected to a WAP that provides internet access. An example topology of this network configuration is demonstrated in Fig. \ref{fig:network-topology}.

\begin{figure}[ht]
    \centering
    \fbox{\includegraphics[width=0.9\linewidth]{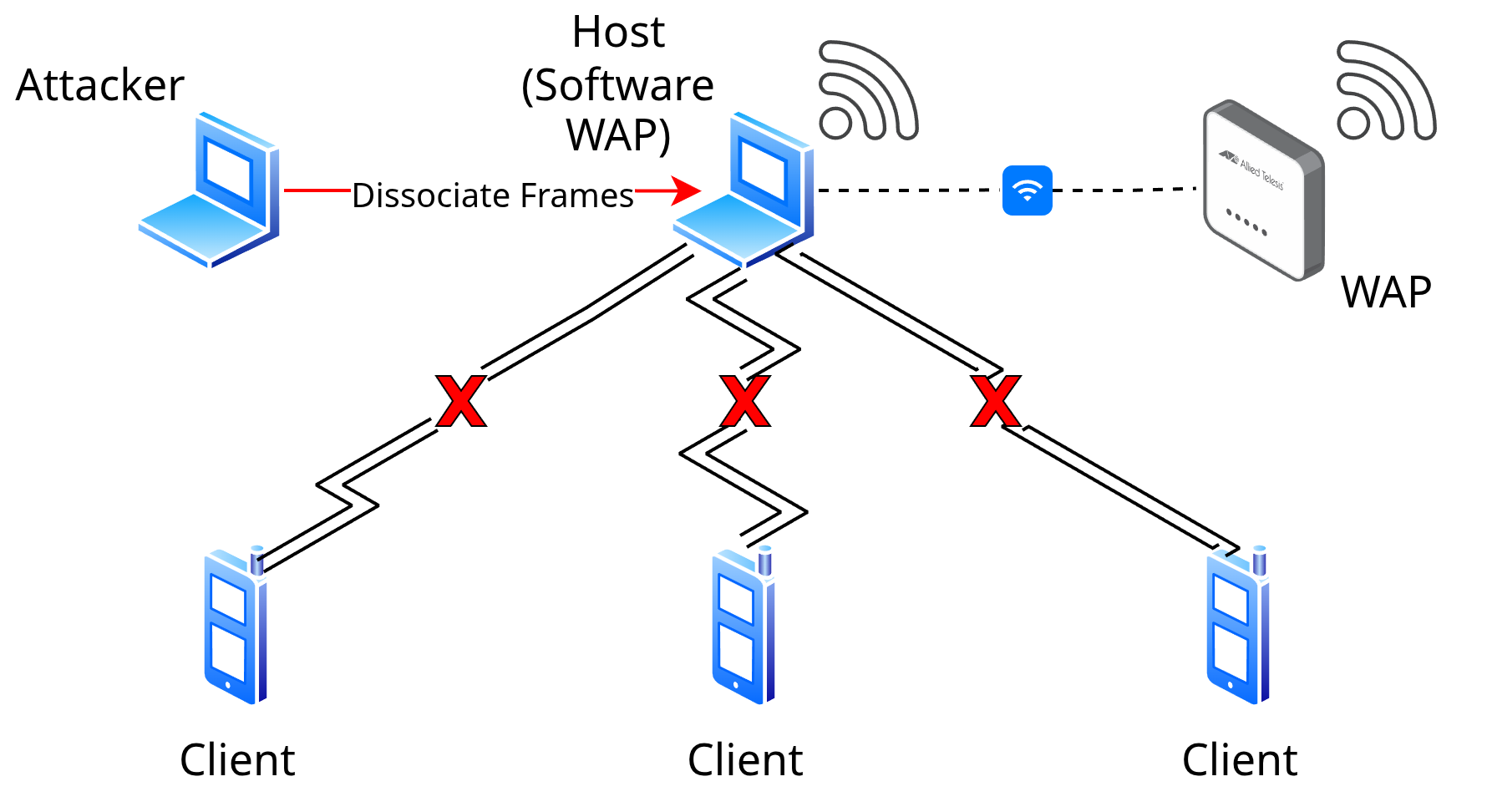}}
    \caption{Deauthentication Attack Network Topology}
    \label{fig:deauthentication-topology}
\end{figure}

The methodologies used include spoofing within a network and forging of dissociate packets to perform the attack. Spoofing is a common attack vector in digital systems that involves attackers pretending to be a trustful entity to gain the trust of the victim \cite{stallings-spoofing}. In the case of a deauthentication attack, the attacker is pretending to be a trusted entity by sending seemingly valid dissociate frames to the software WAP, affecting the clients by forcing them to disconnect from the network provided by the software WAP. Victims include both the software WAP, who mistakenly trusts the dissociate frames, and the clients, who are unjustly disconnected. Fig. \ref{fig:deauthentication-topology} displays an example of a network and its clients facing a deauthentication attack.

Data flow of the simulation involves several components that work in conjunction to ensure the host machine creates a working software WAP that allows for proper connectivity by the clients. The software WAP broadcasts the Service Set Identifier (SSID) of the Wi-Fi network, allowing clients to connect. DHCP and DNS services are set up within the software WAP to automatically assign IP addresses to connected clients and to provide them with internet connectivity. Monitor mode is enabled for the attacker so data frames involving the software WAP and its clients are captured into packet capture (PCAP) files for later inspection. Deauthentication injection occurs through IEEE 802.11 dissociate frames targeting clients connected to the software WAP, forcing disconnections. Disconnection events and any client reconnection attempts are captured as proof of what caused the client to drop connection from the software WAP network.

As a deauthentication attack is a type of DoS, the algorithm used to perform the attack is a brute-force packet injection attack \cite{stallings-brute-force}. A considerable number of packets is sent to the software WAP to ensure that the disruption executes for the clients. Concerning the configuration of the targeted network, networks using no encryption or outdated security protocols are the most straightforward targets \cite{faíscas}. Networks applying modern security standards may still be targeted, though an attacker will find them more difficult as the security measures can serve as a barrier. The difficulty of attacking secure and encrypted networks relies heavily on the type of security protocol being used. For example, WPA3 encrypted networks implement sturdier security and are more difficult to attack than WPA2 networks \cite{sagers-wpa3}.

Table \ref{tab:reproducibility} summarizes the components required to reproduce the experimental setup under the evaluated threat model.

\begin{table}[ht]
  \centering
    \caption{Reproducibility Summary for Wi-Fi Deauthentication Testbed}
    \label{tab:reproducibility}
    \begin{tabular}{p{1.5cm}|p{3cm}|p{3cm}}
      \textbf{Component} & \textbf{Description} & \textbf{References}\\
      \hline
      Topology & Client-host-attacker & Sec. \ref{sec:system-design}, Fig. \ref{fig:network-topology}, Fig. \ref{fig:deauthentication-topology}\\
      \hline
      Source Code & Testbed shell scripts & Sec. \ref{sec:implementation}\\
      \hline
      Tools & Wireshark, aircrack-ng, mdk4, etc. & Sec. \ref{sec:implementation}\\
      \hline
      Data & PCAP files & Sec. \ref{sec:results}\\
      \hline
      Environment & Arch Linux & Sec. \ref{sec:system-design}, Sec. \ref{sec:implementation}\\
      \hline
      Hardware & Wi-Fi devices & Sec. \ref{sec:system-design}\\
    \end{tabular}
\end{table}

\section{Implementation / Experiment Setup}
\label{sec:implementation}

The attack simulation is being performed on two Arch Linux systems on computers with WNICs, so all the tools and methodologies used are based around these platforms. It is important to note that even though Arch Linux is being used, all of the software utilized is available on most other Linux distributions.

Software names correspond to the respective software's name in the Arch Package Database \cite{arch-package}. The following is a list of tools employed for the simulation:
\begin{itemize}
    \item \textbf{aircrack-ng}: Hacking tool focused on the security of Wi-Fi networks. This is the first of the two main tools used in the deauthentication attack. The attacker utilizes it to monitor the software WAP.
    \item \textbf{dnsmasq}: Provides DHCP and DNS services to devices connected to a network. Ensures that the devices are properly configured and can access the internet.
    \item \textbf{hostapd}: Daemon for creating WAPs, including software WAPs. This is how the software WAP the clients connect to in the simulation is created.
    \item \textbf{iproute2}: Utilities for IP routing and managing network interfaces, including WNICs. Useful for viewing information about the networks on the host device used for the simulation.
    \item \textbf{iw}: Command-line tool for configuring wireless devices. Includes many of the commands for manipulating WNICs along with providing supplementary information about WNICs.
    \item \textbf{macchanger}: Utility to manipulate the MAC address of a network device. Used for networking configuration of the software WAP.
    \item \textbf{mdk4}: Wireless attack tool that exploits vulnerabilities in IEEE 802.11 networks. This is the second of the two main tools used in the deauthentication attack. The attacker utilizes it to run a deauthentication attack against the software WAP and its clients.
    \item \textbf{networkmanager}: System service and tools for managing wireless connections. Wireless devices use it to connect to and manage Wi-Fi network connections.
    \item \textbf{nftables}: Framework for packet filtering, NAT, and other tasks involving packet management. The attack simulation utilizes it to perform proper NAT to devices connecting to the software WAP.
    \item \textbf{psmisc}: Miscellaneous tools for the proc file system in Linux. Used in the simulation for restoring normal operation of the host device by killing processes that are no longer needed.
    \item \textbf{ripgrep}: Fast regex search tool. Terminal commands that produce output often employ it to search for relevant information.
    \item \textbf{wavemon}: Terminal tool for monitoring wireless network interfaces. Provides details about wireless networks detected by the host device in a well-formatted interface.
    \item \textbf{wireshark-qt}: Robust tool for analyzing network traffic. Dissociate packets sent in the attack simulation and the subsequent resulting PCAP files can be analyzed with \textit{Wireshark}.
\end{itemize}

Several components of the simulation require configuration. \textit{hostapd}, \textit{dnsmasq}, and \textit{nftables} require heavy configuration as they need to be setup to function as the software WAP that is desired. Custom shell scripts created specifically for this Wi-Fi attack simulation on Arch Linux are available on GitHub\footnote{https://github.com/alex-sir/wifi-deauthentication}.

The test scenario formed for the simulation involves an Arch Linux host system containing a setup of a software WAP and clients with wireless capabilities connecting to the host. The host is connected to a hardware WAP with an internet connection so internet connectivity can be passed to the clients from the host. All networking configuration such as DHCP, DNS, NAT, and port forwarding is handled by the host system.

Experiments with the following Wi-Fi configurations are performed:
\begin{itemize}
    \item Configuration A: Open Wi-Fi network with no security protocol in use
    \item Configuration B: WPA1
    \item Configuration C: WPA2 with PMF turned off
    \item Configuration D: WPA2 with PMF turned on
    \item Configuration E: WPA3
\end{itemize}
Each experiment is repeated five times and is performed with three client devices to simulate the attributes of a real network. The results of the experiments are analyzed and compared with one another.

Performance metrics include the following:
\begin{itemize}
    \item Average percentage of clients connected to the software WAP disconnected by the deauthentication attack.
    \item Median number of dissociate packets until client disconnection occurs.
    \item Median number of seconds until client disconnection occurs.
\end{itemize}
Comparison charts are given that visualize and summarize the results for all of the experiments.

\section{Results}
\label{sec:results}

The Wi-Fi deauthentication attack simulation demonstrates that active security mechanisms for Wi-Fi management frames are highly effective against classical deauthentication attacks. Calculations are performed based on the results of five Wi-Fi configuration experiments  repeated five times with three client devices connected to the host. Data analysis is performed by viewing capture files (.cap) using \textit{Wireshark}. Each deauthentication/dissociate packet is 26 bytes in size. The following filters are used in \textit{Wireshark} to view the relevant network traffic:
\begin{itemize}
    \item Deauthentication/disassociation frames: \begin{verbatim}wlan.fc.type_subtype == 0x0c ||
wlan.fc.type_subtype == 0x0a\end{verbatim}
    \item Deauthentication/disassociation frames for a client: \begin{verbatim}wlan.addr == AA:BB:CC:11:22:33 &&
(wlan.fc.type_subtype == 0x0c ||
wlan.fc.type_subtype == 0x0a)\end{verbatim}
    \item Client addresses (association requests): \begin{verbatim}wlan.fc.type_subtype == 0x0000\end{verbatim}
\end{itemize}

\begin{figure}[ht]
    \centering
    \fbox{\includegraphics[width=0.9\linewidth]{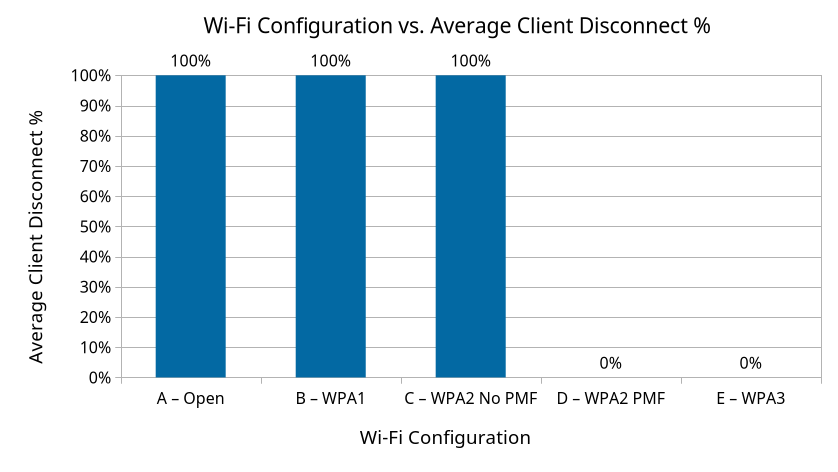}}
    \caption{Wi-Fi Configuration vs. Average Client Disconnect \%}
    \label{fig:disconnect}
\end{figure}

The average for the client disconnect percentage, median number of packets until disconnect, and median time to disconnect in seconds is computed for all five configurations. Median is utilized to provide robustness against outliers that occurred because of traffic on the utilized network. Fig. \ref{fig:disconnect} displays the results of the average disconnect percentage for clients connected to the software WAP in the experiments.

The open, WPA1, and WPA2 with no PMF Wi-Fi configurations resulted in 100\% of clients being disconnected from the software WAP. In comparison, the WPA2 with PMF and WPA3 configurations had 0\% of their clients disconnected. Enabling PMF in WPA2 or using WPA3, which requires PMF to be enabled, significantly increases security for the clients that are connected to the software WAP. Client devices remain highly vulnerable to deauthentication attacks if the network is open, PMF is disabled, or the network uses outdated security protocols such as WPA1.

\begin{figure}[ht]
    \vspace{0.02in}
    \centering
    \fbox{\includegraphics[width=0.9\linewidth]{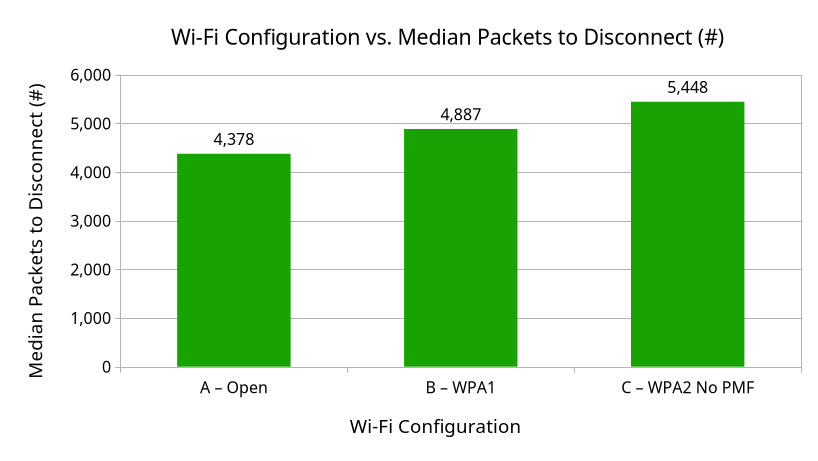}}
    \caption{Wi-Fi Configuration vs. Median Packets to Disconnect (\#)}
    \label{fig:packets}
\end{figure}

Next, is the median number of packets that were sent by the attacker until the clients disconnected. Fig. \ref{fig:packets} shows this data.

WPA2 with PMF and WPA3 are not displayed in this case since clients were not able to be disconnected for those Wi-Fi configurations. Open, WPA1, and WPA2 with no PMF configurations required several thousand deauthentication/dissociate packets to be sent for full deauthentication of the connected client devices. The open configuration required 4,378 packets, WPA1 required 4,887 packets, and WPA2 with no PMF required 5,448 packets. In all three configurations, successful deauthentication required the attacker to transmit many forged management frames in rapid succession. This is consistent with prior descriptions of practical deauthentication attacks that rely on continuous deauthentication frame injection \cite{schepers-ranganathan-vanhoef}.

\begin{figure}[ht]
    \centering
    \fbox{\includegraphics[width=0.9\linewidth]{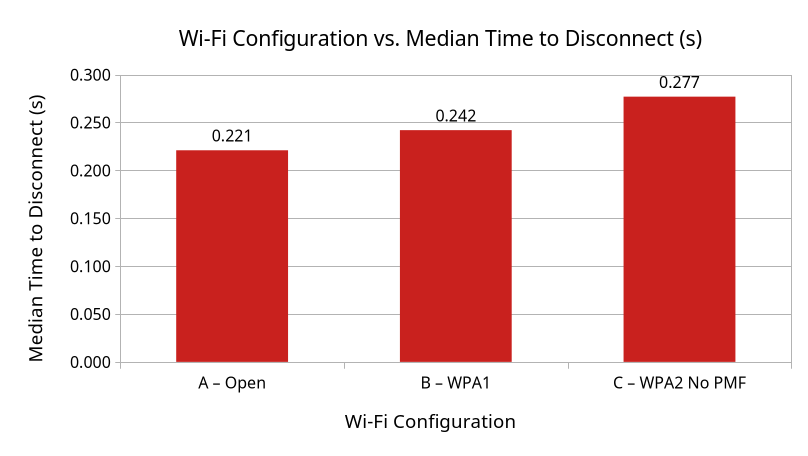}}
    \caption{Wi-Fi Configuration vs. Median Time to Disconnect (s)}
    \label{fig:time}
\end{figure}

Lastly, is the median amount of time in seconds it took for clients to disconnect from the software WAP. This is displayed in Fig. \ref{fig:time}.

Once again, WPA2 with PMF and WPA3 are not relevant due to them being successful in protecting against the deauthentication attacks. The open configuration took 0.221 seconds to disconnect clients, WPA1 took 0.242 seconds, and WPA2 with no PMF took 0.277 seconds. This demonstrates quick disruption of network connectivity for clients once sustained deauthentication traffic begins for these Wi-Fi configurations.

\section{Discussion}
\label{sec:discussion}

The results of the Wi-Fi deauthentication attack experiments demonstrate the importance of utilizing the proper wireless security protocols for wireless networks. Setting a wireless network to not use a security protocol, using an outdated security protocol such as WPA1, or not enabling PMF with WPA2 leaves the network highly vulnerable to deauthentication attacks from cyber attackers. A complete DoS is possible when not using wireless security protocols that implement PMF such as WPA2 with PMF or WPA3. The number of packets and the amount of time it takes to attack vulnerable WAPs, whether with encryption or not, is low. Any malicious actor with the desire to cause a DoS to a wireless network with a vulnerable configuration has a streamlined method to perform the attack, so network administrators are advised to exercise caution when configuring wireless networks for others to use. Users of these networks must also be vigilant and ensure that the networks they connect to are not straightforward targets for attackers.

These types of DoS attacks are possible because the standard that defines deauthentication and disassociation frames in IEEE 802.11 allows for these frames to be sent in an unencrypted manner \cite{schepers-ranganathan-vanhoef}. This is not out of negligence, but it is purposely done for ease of use and to enable recovery mechanisms. Although the experiments performed in this paper did not result in deauthentication of clients in WPA3 networks, with more robust attack systems and a deeper understanding of wireless network security protocols it is possible to perform DoS attacks on wireless networks utilizing the most modern security protocols \cite{chatzoglou-kambourakis-kolias}.

This work is related to modern cybersecurity due to the proliferation of wireless networks in day-to-day life \cite{rikitianskaia}. Users of these networks and network administrators do not want to become victims of these attacks, as not being able to use the network when desired is bothersome. Whether in a public area or in a private residence, these wireless networks ultimately rely on the same set of standards set by the IEEE.

Averting Wi-Fi deauthentication attacks is difficult due to the fact that security measures implemented to avert these attacks depend on many facets to determine their effectiveness. For example:
\begin{itemize}
    \item Do the client devices have support for the latest security standards?
    \item Do the Wi-Fi routers and WAPs have support for the latest security standards?
    \item Can vendor patches or updates be applied to existing devices?
    \item Difficulty and length of time to thoroughly implement new security standards.
\end{itemize}
With that said, there are proposals for preventing Wi-Fi DoS attacks from occurring in the future. One possibility is to increase the robustness of PMF through methods such as requiring the silent discarding of invalid frames \cite{schepers-ranganathan-vanhoef}. Another possibility is to attach a timeout to the handshake process performed between a WAP and a client device, effectively shutting down a DoS attack vector \cite{schepers-ranganathan-vanhoef}.

\section{Limitations \& Next Steps}
\label{sec:limitations}

This study performs experiments, analyses the effects, and discusses Wi-Fi deauthentication attacks on networks implementing certain wireless security protocols. The most significant limitation of this study is the access to hardware to perform the experiments. Hardware is strictly limited to what is already available, which is a small number of client devices and a Wi-Fi 6 WAP. Ideally, a greater number of client devices are utilized (20+) to reflect real-world environments more precisely. A more modern WAP would allow for robust testing of the latest wireless security protocols such as Wi-Fi 7. Additionally, more effective software for performing the deauthentication attacks could yield vastly different results, such as WPA2 with PMF and WPA3 encrypted networks being successfully breached.

Future work will involve a stricter focus on the latest security protocols (e.g., WPA3) as the use of them in wireless networks becomes more common. Another area of focus will be on evaluating how effective the security is for each Wi‑Fi generation (e.g., Wi‑Fi 5, Wi‑Fi 6, and beyond). Though newer Wi-Fi generations are closely related to superior security protocols, it is still a topic of interest due to factors such as backwards compatibility \cite{ieee-wifi-evolution}.

\section{Conclusion}
\label{sec:conclusion}

In summary, this work demonstrates how deauthentication and disassociation frames can be exploited by threat actors to perform denial-of-service attacks against misconfigured and legacy Wi-Fi networks. Our results indicate that open, WPA1, and WPA2 with no PMF network deployments remain highly vulnerable to deauthentication injection. In contrast, modern configurations that enforce PMF, such as WPA2 with PMF and WPA3, exhibit strong resilience under the tested conditions. These findings confirm that encryption alone is insufficient without management-frame protection. Furthermore, it highlights that improper configuration of wireless networks remains a major operational security risk. As a consequence, network administrators and users must prioritize the use of PMF-capable security modes to mitigate denial-of-service threats in contemporary Wi-Fi deployments.

\end{document}